\documentstyle[twoside,fleqn,npb,epsfig]{article}
\newcommand{\AmS}{{\protect\the\textfont2
  A\kern-.1667em\lower.5ex\hbox{M}\kern-.125emS}}

\def\NP{Nucl. Phys. B}
\def\PRL{Phys. Rev. Lett.}
\def\PRD{Phys. Rev. D}
 
\def\ppbar{$p\overline{p}~$}             
\def\pbarp{$\overline{p}p~$}             
\def\pt{$p_T~$}                          
\def\met{\mbox{${\hbox{$E$\kern-0.6em\lower-.1ex\hbox{/}}}_T~$}} 
\def\ipb{pb$^{-1}~$}                     
\def\mw{$M_W~$}                          
\def\mz{$M_Z~$}                          
\def\D0{D\O}                            
\def\etal{{\sl et al.}}                 
\newcommand{\Wev}{$W\rightarrow e\nu$}

\def\Zee{\mbox{$Z \rightarrow e e \ $}}

\def\pbarp{\mbox{$\overline{p}p$}}

\def\pt{\mbox{$p_T$}}

\def\mz{\mbox{$m_Z$}}
\def\mw{\mbox{$m_W$}}
\def\gw{\mbox{$\Gamma_W$}}

\def\d2sg{\mbox{$d^2_{\sigma}$}}

\def\Wtv{\mbox{$W\to \tau\nu$}}
\def\Ztt{\mbox{$Z\to \tau \tau$}}

\def\ipb{pb$^{-1}$}
\def\ptmiss{\mbox{${\not{\hbox{\kern-2.3pt$p_T$}}}$}}
\def\vptmiss{\mbox{${\not{\hbox{\kern-2.3pt$\vec{p}_T$}}}$}}
\catcode`@=11 
\def \gsim{\mathrel{\mathpalette\@versim>}}
\def \lsim{\mathrel{\mathpalette\@versim<}}
\def \@versim#1#2{\lower0.4ex\vbox{\baselineskip\z@skip\lineskip\z@skip
     \lineskiplimit\z@\ialign{$\m@th#1\hfil##\hfil$%
     \crcr#2\crcr\sim\crcr}}}
\catcode`@=12 
\def\etal{{\it et al.}}
\newcommand{\wxsec}{2310}

\newcommand{\wnum}{67078}
\newcommand{\wacc}{0.465}
\newcommand{\waccerr}{0.004}
\newcommand{\weff}{0.671}
\newcommand{\wefferr}{0.009}
\newcommand{\wfqcd}{0.064}
\newcommand{\wfqcderr}{0.014}

\newcommand{\wacczinw}{0.133}
\newcommand{\wacczinwerr}{0.034}
\newcommand{\wacctau}{0.021}
\newcommand{\wacctauerr}{0.002}

\newcommand{\zxsec}{221}

\newcommand{\znum}{5397}
\newcommand{\zacc}{0.366}
\newcommand{\zaccerr}{0.003}
\newcommand{\zeff}{0.744}
\newcommand{\zefferr}{0.011}
\newcommand{\zfdy}{0.012}
\newcommand{\zfdyerr}{0.001}
\newcommand{\zfqcd}{0.045}
\newcommand{\zfqcderr}{0.005}

\newcommand{\lumb}{84.5}
\newcommand{\lumberr}{3.6}

\newcommand{\rxsec}{10.43}

\newcommand{\rnum}{12.43}
\newcommand{\rnumerr}{0.18}
\newcommand{\racc}{0.787}
\newcommand{\raccerr}{0.007}
\newcommand{\reff}{1.108}
\newcommand{\refferr}{0.007}

\newcommand{\brwev}{0.1066}
\newcommand{\brstat}{0.0015}
\newcommand{\brsyst}{0.0021}
\newcommand{\brthy}{0.0011}
\newcommand{\brnlo}{0.0011}

\newcommand{\gwstat}{0.030}
\newcommand{\gwsyst}{0.041}
\newcommand{\gwthy}{0.022}
\newcommand{\gwnlo}{0.021}

\newcommand{\gwinvgev}{0.168}

\title{Recent Results from the Tevatron Fixed Target and Collider Experiments}

\author{Cecilia E.\ Gerber\address{Fermi National Accelerator Laboratory\\
P.O. Box 500, Batavia, IL 60510, USA}
}
\begin{document}
\lefthyphenmin=2
\righthyphenmin=3

\begin{abstract}
We present a review of recent QCD related results from the Fermilab 
Tevatron fixed target and collider experiments. Topics include jet and
boson production, $W$ boson and top quark mass measurements, and 
studies of $CP$ violation.
\end{abstract}

\maketitle

\section{INTRODUCTION}

Quantum Chromodynamics (QCD) emerged as a mathematically consistent theory in
the 1970s, and nowadays is regarded as one of the cornerstones of the
Standard Model. One of the triumphs of modern particle physics has been the
extent to which QCD has successfully accounted for the strong interaction 
processes observed experimentally at hadron colliders. 
Some of the processes studied include hadronic jet, heavy quark, and gauge 
boson production.

The number of new results from the Fermilab Tevatron accelerator that
are being presented in over twenty parallel sessions at this conference is 
overwhelming. The two collider detectors, CDF and \D0, have finished taking 
data in 1996; new results on Jet and Boson properties are based on these
large data sets of $\sim 100\;\rm pb^{-1}$ integrated luminosity. Both
collaborations are upgrading their detectors in preparation for Run II,
scheduled to start in the year 2000. Results from the fixed target experiments
are based on data taken during the last fixed target run that ended 
in 1997. The upcoming fixed target run is scheduled for later this year.

In this summary we review QCD results that are new since last year and that 
are presented in greater detail in the parallel sessions. We also include the
new Tevatron results on the mass of the $W$ boson and the top quark from CDF
and \D0. These measurements are used to constrain the mass of the Higgs boson.
In addition, we present recent studies of $CP$ violation. The KTeV 
collaboration has clearly 
observed direct $CP$ violation in the Kaon system. CDF
observes the first indication of $CP$ violation in the $b$ quark system.
New results from the NuTeV collaboration are being presented in a different
summary contribution to this conference~\cite{jorge}.

\section{JET PRODUCTION IN PROTON--\\ANTIPROTON COLLISIONS}

At the Tevatron energies, the dominant process in \ppbar collisions is jet 
production. Within the framework of QCD, inelastic 
scattering between a proton and an antiproton can be described as an elastic 
collision between a single proton constituent and a single antiproton 
constituent. These constituents are called partons. After the 
collision, the outgoing partons manifest themselves as localized streams of 
particles referred to as ``jets''. Theoretical predictions for jet production 
are given by the folding of the parton scattering cross sections with 
experimentally determined parton density functions (pdf's). These 
predictions have recently improved with next--to--leading order (NLO) QCD 
scattering calculations~\cite{EKStheory,Aversatheory,GGKtheory} and new, 
accurately measured 
pdf's~\cite{jet4,jet5}. Some of the questions that can be addressed with 
studies of jet production are testing of NLO QCD, extraction of pdf's,
measuring the value of the strong coupling constant $\alpha_s$, and testing
quark compositeness.  

\subsection{Inclusive Jet Cross Section}

\begin{figure}[htb]
\vspace{9pt}
\hskip0.4cm
\psfig{figure=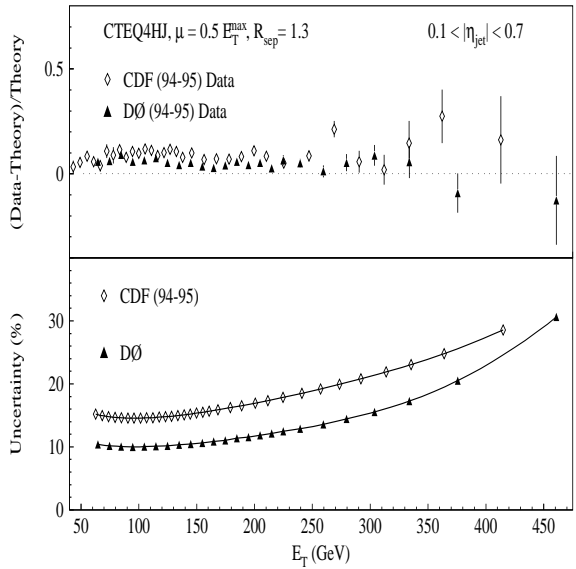,height=7.6cm,width=6.8cm}
\caption{Inclusive jet cross section in the central rapidity
region for CDF and \D0, plotted versus jet $E_{T}$. The data points are
shown with statistical uncertainties.
The systematic uncertainty on the ratio is shown in the bottom half of the
plot.}
\label{fig:central_jets}
\end{figure}

The \D0 and CDF  collaborations
measure the central inclusive jet 
cross section  in \ppbar collisions at $\sqrt{s}=1.8\;\rm TeV$ using an 
integrated luminosity of $92\;\rm pb^{-1}$ and $87\;\rm pb^{-1}$, 
respectively. The inclusive double differential jet cross section can
be expressed as:
\[ 
d^2\sigma/(dE_T d\eta)=(N_{Jet})/(\varepsilon\Delta E_T\Delta\eta\int L dt)\]
where $N_{Jet}$ is the total number of jets observed in a certain
 jet transverse energy $E_T$ bin, $\varepsilon$ is the selection efficiency, 
$\Delta E_T$ is the bin width, $\Delta\eta$ is the
pseudorapidity range considered, and $\int L dt$ is the integrated luminosity
associated with the data set.
The cross sections are measured in the pseudorapidity interval
$0.1<|\eta|<0.7$ (CDF,~\cite{akopian}), 
and the two pseudorapidity ranges 
$|\eta|<0.5$ and $0.1<|\eta|<0.7$ (\D0,~\cite{d0incjetprl}). 
Figure~\ref{fig:central_jets} shows the ratio plot (Data-Theory)/Theory for the
$0.1<|\eta|<0.7$ rapidity range for CDF and \D0 data compared to NLO QCD.

\begin{figure}[htb]
\vskip-0.8cm
\hskip-1.0cm
\vspace{9pt}
\psfig{figure=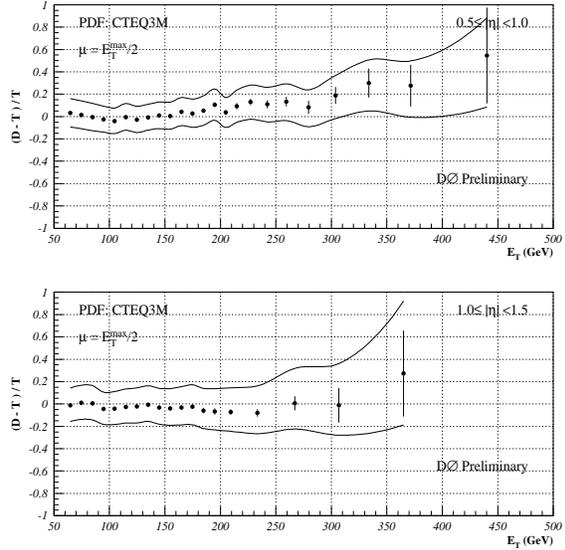,height=8.5cm,width=8.3cm}
\caption{Rapidity dependence of the inclusive jet cross section plotted
versus jet $E_{T}$ from \D0.
The systematic uncertainty on the ratio is shown as a band.}
\label{fig:forward_jets}
\end{figure}

In addition, \D0 presented for the first time the preliminary~\cite{daniel}
measurement of the rapidity dependence of the inclusive jet cross section, 
which extends the measurement to two forward rapidity regions:
$0.5<|\eta|<1.0$ and $1.0<|\eta|<1.5$. 
Figure~\ref{fig:forward_jets} shows the ratio plot (Data-Theory)/Theory for
this measurement. 
All the measurements show good agreement with the NLO QCD 
predictions currently available.

\begin{figure}[t]
\vspace{9pt}
\psfig{figure=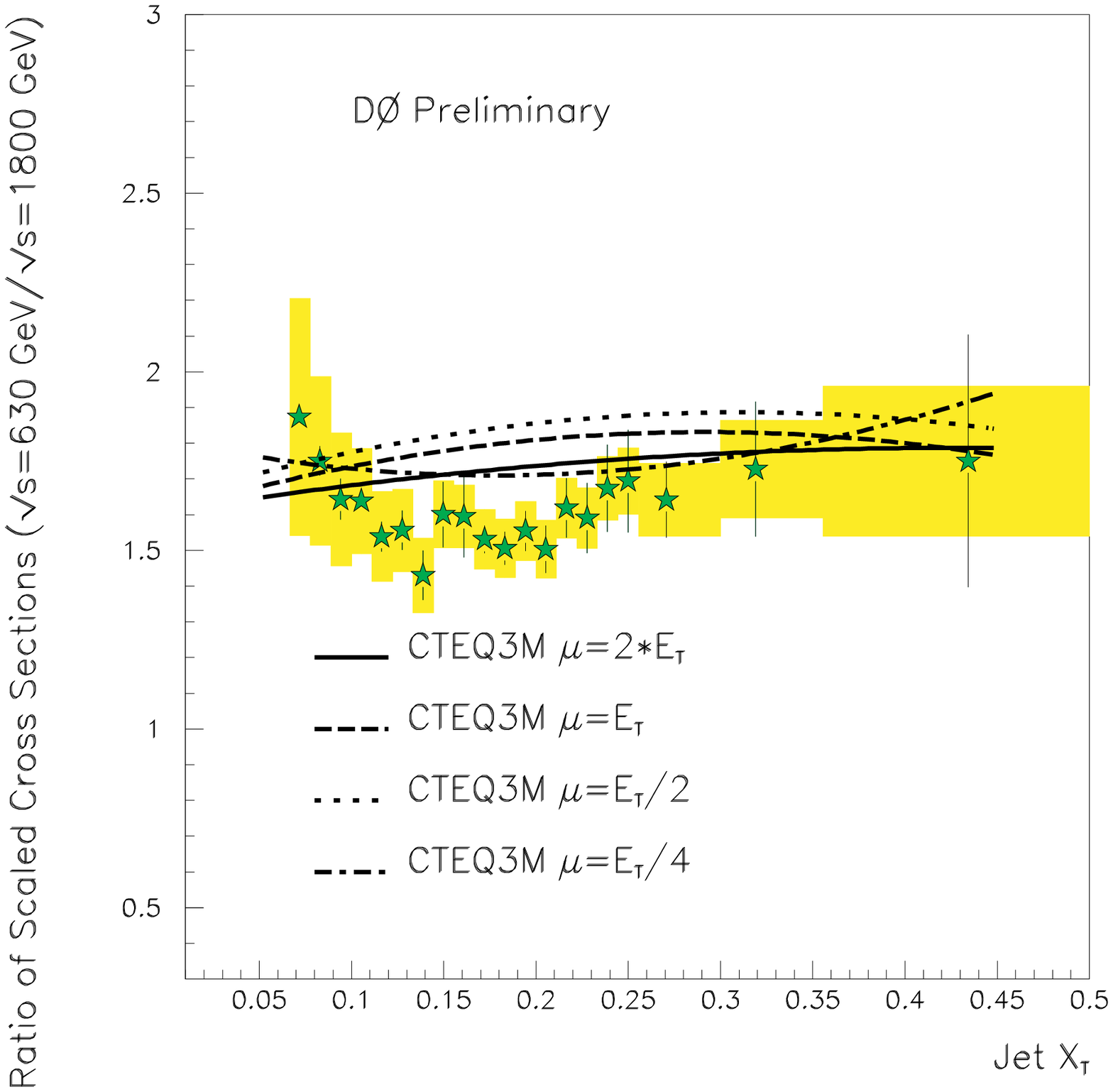,height=7.6cm,width=7.3cm}
\caption{Scale invariant cross section from \D0. Data points are shown with 
statistical uncertainty; systematic uncertainty is shown as a band. The 
NLO QCD theoretical predictions for different renormalization scales are shown
as lines.}
\label{fig:scale_inv_d0}
\end{figure}

\begin{figure}[t]
\vskip-0.8cm
\vspace{9pt}
\psfig{figure=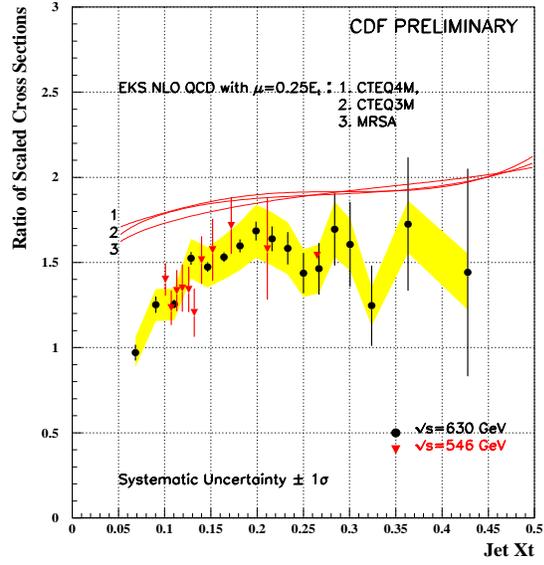,height=8.5cm,width=7.9cm}
\caption{Scale invariant cross section from CDF. Data points are shown with 
statistical uncertainty; systematic uncertainty is shown as a band. The 
NLO QCD theoretical predictions for different renormalization scales are shown
as lines.}
\label{fig:scale_inv_cdf}
\end{figure}

Although the Tevatron nominally operated at a center of mass energy of
$1.8\;\rm TeV$, a short period of the time was devoted to
collect data at the lower center of mass energy of $\sqrt{s}=630\;\rm GeV$.
\D0~\cite{daniel} and CDF~\cite{akopian} 
measure the ratio of scale invariant cross section
$\sigma_S=(E_T^3/2\pi)(d^2\sigma/dE_T d\eta)$ at two center of mass
energies as a function of Jet $x_T=E_T/(\frac{\sqrt{s}}{2})$.
Figures~\ref{fig:scale_inv_d0} and \ref{fig:scale_inv_cdf} show the preliminary
results for
\D0 and CDF respectively. NLO QCD overestimates the \D0 data by almost three
standard deviations in the medium range of $x_T$. 
The disagreement between data and theory 
is even worse for the CDF data at low $x_T$. 
A good quantitative agreement between \D0 data and NLO QCD can be obtained
if different renormalization scales are used in the theoretical calculation
at the two different center--of--mass energies. For instance,  a scale
of $\mu=2 E_T$ at $\sqrt{s}=630\;\rm GeV$ and of 
 $\mu=E_T/2$ at $\sqrt{s}=1800\;\rm GeV$ reproduces the \D0 data best.

CDF~\cite{akopian} 
extracts the value of the strong coupling constant $\alpha_s$ in the jet
$E_T$ range from $40-250\;\rm GeV$ by comparing the measured inclusive jet
cross section to the NLO  {\sc JETRAD}~\cite{jetrad} Monte Carlo. 
The evolution of the coupling
constant over a wide range of scales is clearly observed 
and is in agreement with
QCD predictions. The measured $\alpha_s(E_T)$ is evolved to 
 $\alpha_s(M_Z)$ using a two--loop renormalization group equation. The 
preliminary measurement is $\alpha_s(M_Z)=0.1129\pm0.0001{\rm (stat)}
^{+0.0078}_{-0.0089}{\rm (exp\;syst)}$. 

\subsection{Dijet Production}

\begin{figure}[htb]
\vspace{9pt}
\vskip-0.5cm
\hskip-0.61cm
\psfig{figure=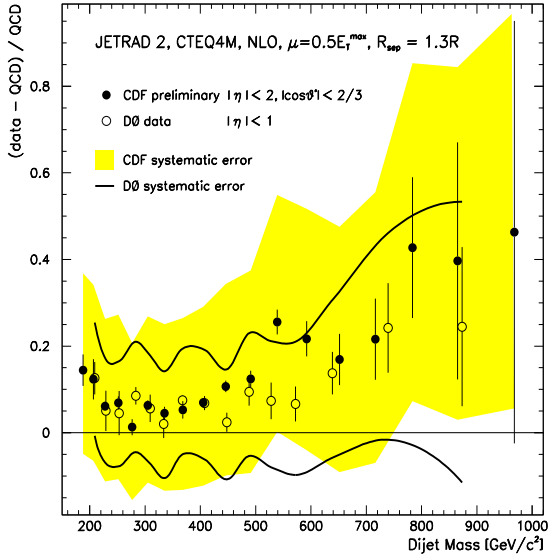,height=8.5cm,width=8.0cm}
\caption{Dijet cross section as a function of the dijet invariant mass from
CDF and \D0. Data are compared to NLO QCD with CTEQ4M parton density function.}
\label{fig:dijetmass}
\end{figure}

CDF and \D0 measure the jet production cross section for events with two
jets as a function of the dijet invariant 
mass. The results are shown in figure~\ref{fig:dijetmass}. NLO QCD is in good 
agreement with the data. \D0's measurement of the dijet mass 
spectrum~\cite{d0dijetmass} 
is used to search for quark compositeness, which would manifest 
itself as an excess of events at high masses. 
A mass scale $\Lambda$ characterizes both
 the strength of the quark substructure binding and the physical size of the 
composite state. Limits are set assuming that $\Lambda\gg\sqrt{s}$ such that 
quarks appear to be point-like and the coupling can be approximated by a 
four--Fermion contact interaction. The best sensitivity is obtained by taking
the ratio of the dijet cross sections for events in which both jets are
central ($|\eta|<0.5$) and events in which both jets are forward 
($0.5<|\eta|<1.0$). The $95\%$ confidence level lower limits on the mass scale
are $\Lambda^+=2.7\;\rm TeV$ and $\Lambda^-=2.4\;\rm TeV$ for destructive and
constructive interference models respectively.

\begin{figure}[htb]
\vspace{9pt}
\psfig{figure=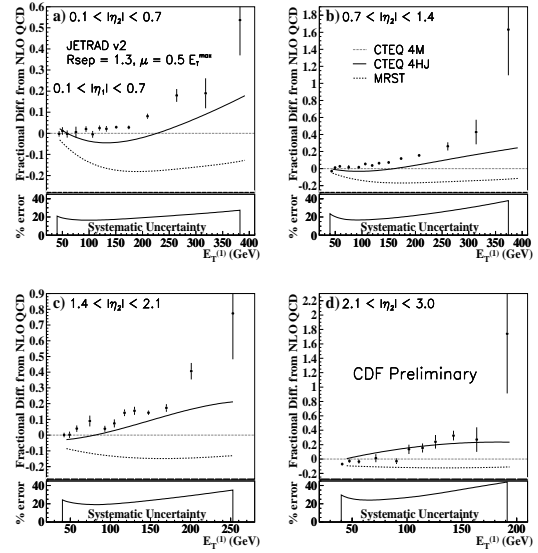,height=7.6cm,width=7.3cm}
\caption{Inclusive Dijet differential cross section as a function of the 
leading central jet $E_T$ from CDF. Data are compared to the prediction by
{\sc JETRAD} using different parton density functions.}
\label{fig:cdfdijet}
\end{figure}

CDF measures the inclusive dijet differential cross section~\cite{frank}
$d^3\sigma/(dE^T_1 d\eta_1 d\eta_2)$ as a function of the trigger jet
$E^T_1$. The trigger jet is central ($0.1<|\eta|<0.7$); the second jet
pseudorapidity is in one of the following four bins:
$0.1<|\eta|<0.7$, $0.7<|\eta|<1.4$, $1.4<|\eta|<2.1$, $2.1<|\eta|<3.0$.
The four resulting cross sections are shown in figure~\ref{fig:cdfdijet}.
The measurement is sensitive to the choice of pdf, and CTEQ4HJ qualitatively 
reproduces CDF data best.
 
\begin{figure}[htb]
\vspace{9pt}
\hskip-1.25cm
\psfig{figure=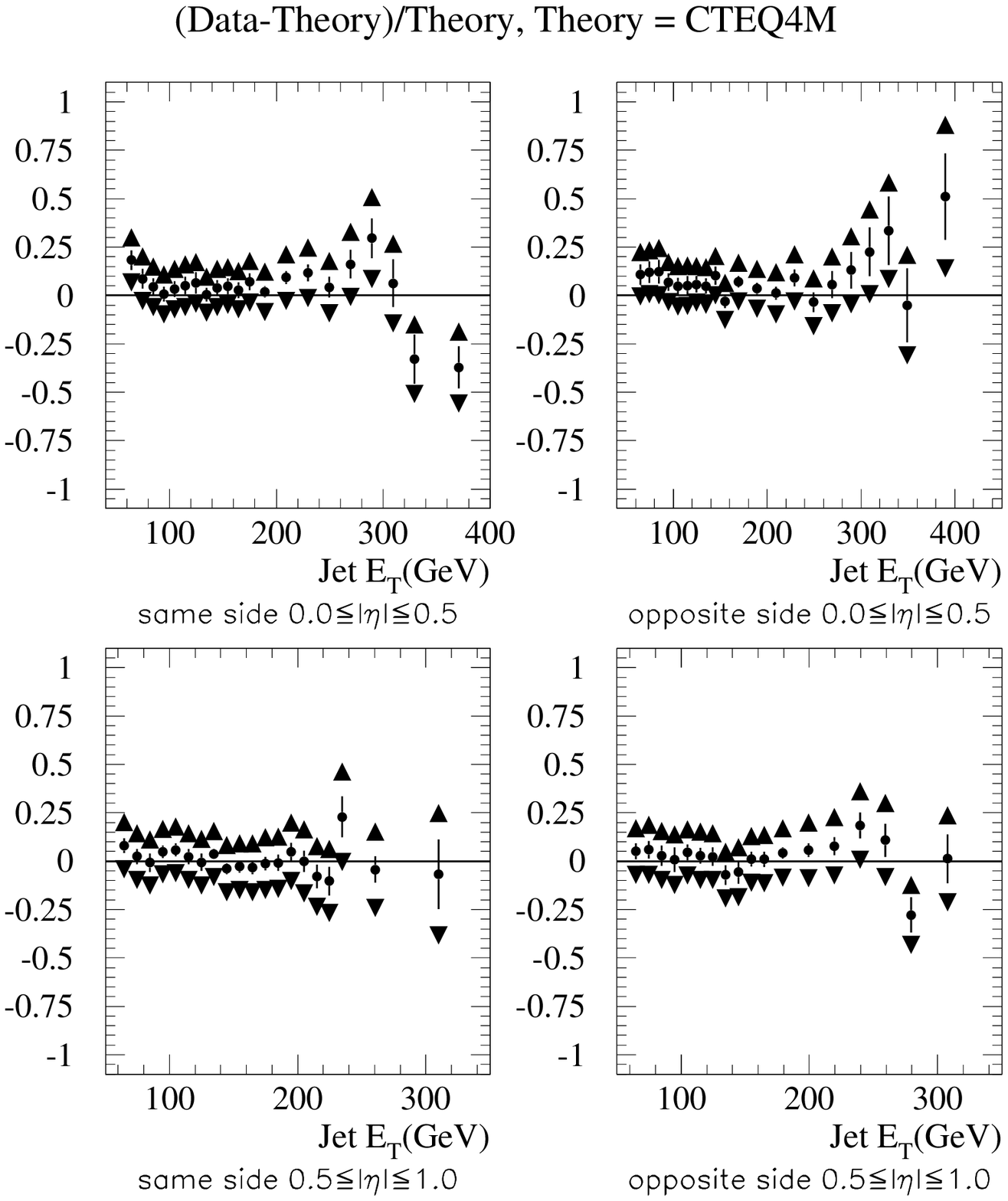,height=8.5cm,width=8.5cm}
\vskip-0.5cm
\caption{Inclusive Dijet Differential cross section from \D0 for
``same side''(left) and ``opposite side''(right) jet topologies. Data are
compared to NLO QCD with the CTEQ4M parton density function.}
\label{fig:d0dijet1}
\end{figure}

\begin{figure}[htb]
\vspace{9pt}
\hskip-1.25cm
\psfig{figure=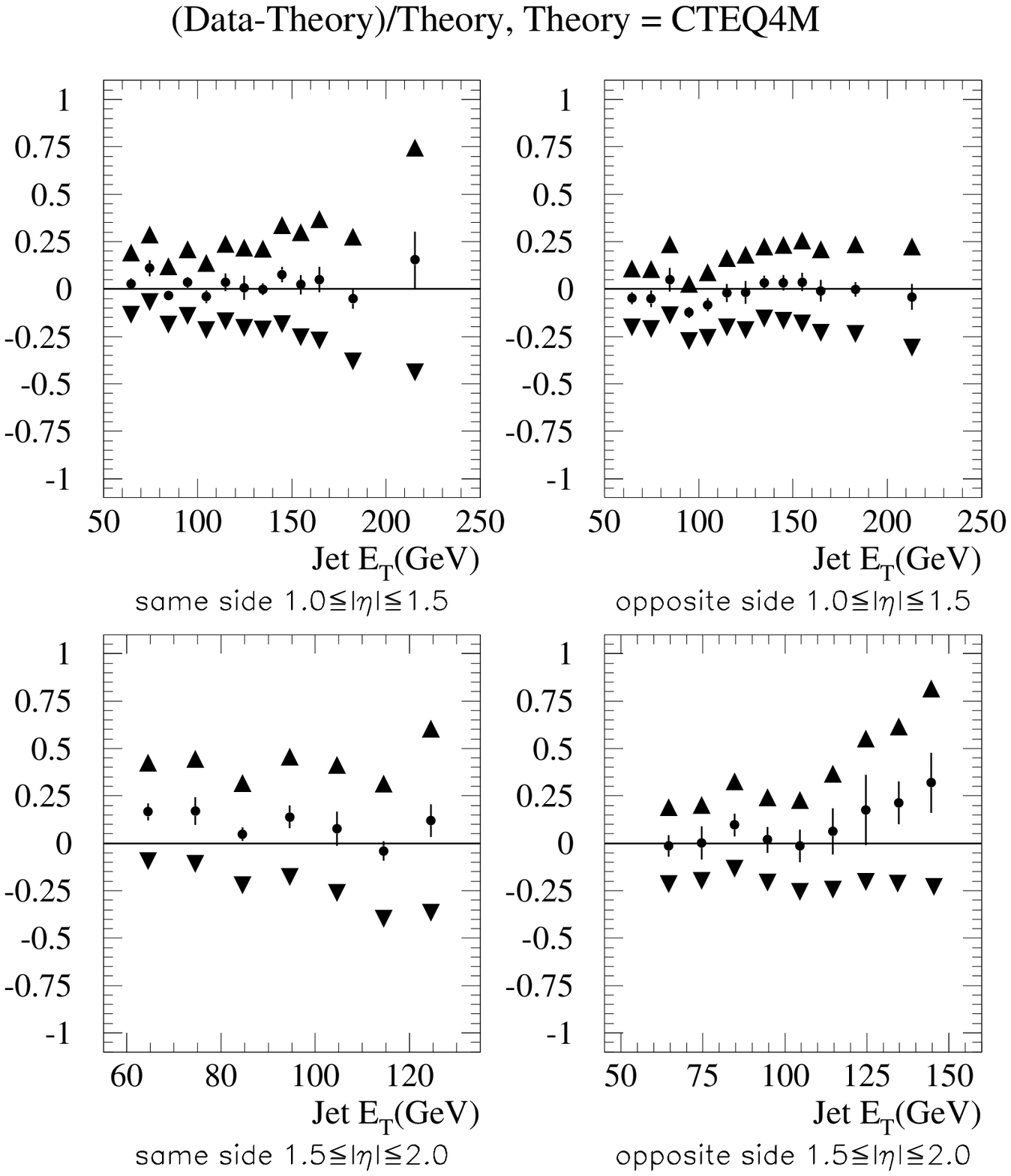,height=8.5cm,width=8.5cm}
\vskip-0.5cm
\caption{Inclusive Dijet Differential cross section from \D0, for
``same side''(left) and ``opposite side''(right) jet topologies. Data are
compared to NLO QCD with the CTEQ4M parton density function.}
\label{fig:d0dijet2}
\end{figure}

The \D0 calorimeter allows the measurement of the energies 
of jets in the very forward rapidity 
region permitting a determination of the inclusive dijet differential cross 
section~\cite{heidi} 
as a function of the leading and next--to--leading jet $E_T$ in four
pseudorapidity bins: $|\eta|<0.5$, $0.5<|\eta|<1.0$, $1.0<|\eta|<1.5$, 
$1.5<|\eta|<2.5$. Two topologies are considered: $\eta_1=\eta_2$ 
(``same side'') and $\eta_1=-\eta_2$ (``opposite side''), with both
jets required to be in the same $|\eta|$ bin.   The eight resulting
cross sections are shown in figures~\ref{fig:d0dijet1}~and~\ref{fig:d0dijet2}.
The measurement is 
sensitive to the choice of pdf and CTEQ4M qualitatively 
reproduces the \D0 data well. 
  
\subsection{Subjet Multiplicity in Quark and Gluon Jets}

\D0 measures the subjet multiplicity in jets reconstructed using the
$k_T$ algorithm~\cite{rob}. Jets with
$55<E_T<100\;\rm GeV$ and $|\eta|<0.5$ are selected from data taken
at two center--of--mass energies,
$\sqrt{s}=1800\;\rm GeV$ and $\sqrt{s}=630\;\rm GeV$.

\begin{figure}[t]
\vspace{9pt}
\psfig{figure=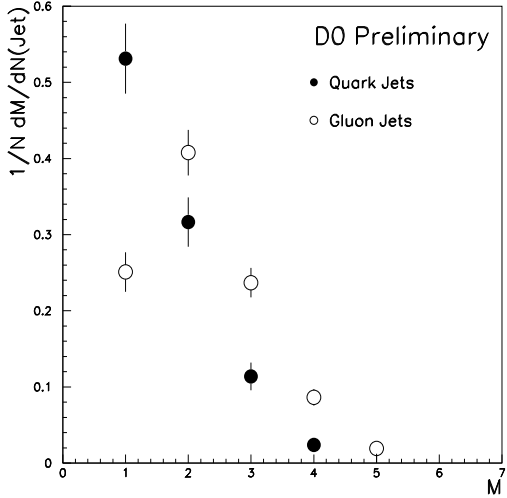,height=7.6cm,width=7.3cm}
\caption{Subjet multiplicity for quark and gluon jets as measured by \D0.}
\label{fig:kt}
\end{figure}

The {\sc HERWIG}~\cite{herwig} Monte Carlo event generator predicts that
$59\%$ of the jets are gluon jets at $\sqrt{s}=1800\;\rm GeV$, 
and $33\%$ of the jets are gluon jets at $\sqrt{s}=630\;\rm GeV$. This
information is used as input to the analysis to extract the
average subjet multiplicity in gluon ($<N_G>$) and quark ($<N_Q>$)
jets. \D0 clearly distinguishes, on a statistical bases, 
between quark and gluon jets, as can be seen 
in figure~\ref{fig:kt}. The measured value of
$R\equiv(<N_G> -1)/(<N_Q> -1)=1.91\pm0.04{\rm (stat)})
^{+0.23}_{-0.19}{\rm (syst)}$ agrees 
with the Monte Carlo prediction of $R=1.86\pm0.08{\rm (stat)}$. 

\subsection{Diffractive Jet Production}

\begin{figure}[t]
\vspace{9pt}
\vskip-1.2cm
\psfig{figure=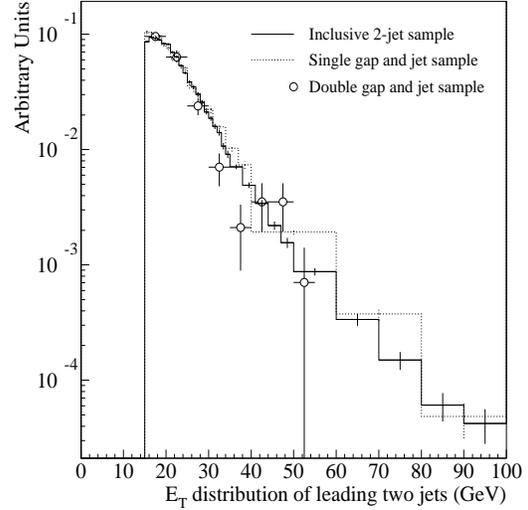,height=7.6cm,width=7.3cm}
\vskip-1.0cm
\caption{$E_T$ distributions of the leading two jets for three
data samples at $\sqrt{s}=1800\;\rm GeV$ from \D0: inclusive
dijet sample (solid histogram), sample with one forward
rapidity gap (dotted line) and double gap events (circles).}
\label{fig:d0diff1}
\end{figure}

\D0 observes the production of dijet events produced in conjunction with
two forward rapidity gaps (no calorimeter or scintillator hits in
$3.0<|\eta|<5.2$) along the directions of each of the initial beam
particles in proton--antiproton collisions at the center of mass energies of
 $\sqrt{s}=1800\;\rm GeV$ and $\sqrt{s}=630\;\rm GeV$~\cite{kristal}.
 This topology is  consistent with Hard Double Pomeron exchange. It is
interesting to examine the $E_T$ spectrum of the jets produced in these
diffractive events. Figure~\ref{fig:d0diff1} shows the $E_T$ spectra of the two
leading jets for an inclusive sample with two central jets greater than 
$15\;\rm GeV$ (solid histogram), a sample with the additional requirement of 
a single forward rapidity gap (dashed histogram) and a sample of double
gap events (open circles). All three spectra are in good agreement where the
data are available, implying that the dynamics of leading jets produced in the
 rapidity gap events appear to be similar to those of inclusive QCD production.
The same behavior is observed in data taken at  
the lower center of mass energy of $\sqrt{s}=630\;\rm GeV$. The $E_T$ spectra
at both center of mass energies looks harder than allowed by the 
$5\%$ rule--of--thumb for the pomeron momentum fraction. 

\begin{figure}[b]
\vspace{9pt}
\vskip-2cm
\psfig{figure=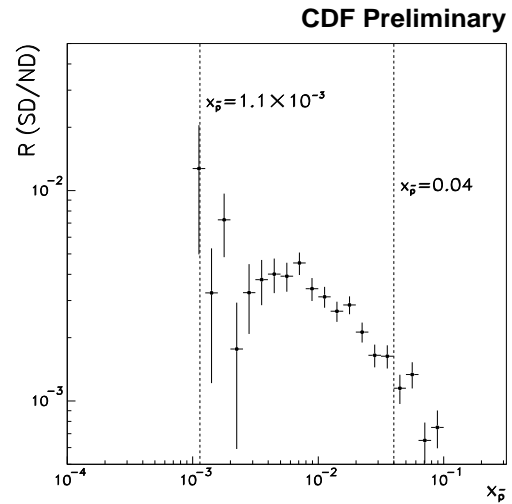,height=7.0cm,width=7.3cm}
\caption{Ratio of single diffractive to non--diffractive dijet events as a 
function of $x_{\overline{p}}$ at $\sqrt{s}=630\;\rm GeV$ from CDF.}
\vskip-2cm
\label{fig:cdfdiff2}
\end{figure}

CDF observes diffractive dijet production associated with a leading 
antiproton in \ppbar collisions at $\sqrt{s}=630\;\rm GeV$
in data taken with the roman--pot trigger~\cite{kerstin}. 
Using the diffractive dijet events in the kinematic region of the momentum 
loss fraction of the
antiproton $0.04 < \xi < 0.10$ and the four momentum transfer 
squared $|t| < 0.2\;\rm GeV^2$,
CDF finds that the cross section ratio of diffractive to non-diffractive
 dijet events as a function of the
momentum fraction of the parton in the antiproton participating 
in the dijet production $x_{\overline{p}}$
decreases with increasing $x_{\overline{p}}$, as can be seen in
figure~\ref{fig:cdfdiff2}. Similar results are observed in data
taken at the center of mass energy of $\sqrt{s}=1800\;\rm GeV$.

\clearpage
\section{BOSON PRODUCTION}

$W$ and $Z$ bosons, the carriers of the weak force, are directly produced in
high energy \ppbar  collisions at the Fermilab Tevatron. 
In addition to probing electroweak physics, the study of the
production of $W$ and $Z$ bosons
provides an avenue to explore QCD, the theory of strong
interactions.  Direct production of photons is also a powerful tool for testing
QCD predictions with fewer 
of the ambiguities associated with jet production and
fragmentation. The measurement of the high mass Drell--Yan cross section above 
the $Z$ mass tests for quark--lepton compositeness.

\subsection{$W$ and $Z$ Production in \ppbar collisions}

Large numbers of $W$ bosons have been
detected by the two collider detectors (CDF and \D0) during the 1992--1996
running period. These samples complement the detailed
studies carried out on the $Z$ boson at LEP and SLC, and also the new $W$
studies from LEP II.

\D0 measures the production cross section times branching ratio for $W$ and $Z$
bosons. 
The product of the $W$ boson cross section and the
branching fraction for \Wev\ is calculated using the relation
\[
\sigma(p \overline{p} \rightarrow W + X) \cdot B(W \rightarrow e \nu)=\]
\[\frac{N_{obs}^W  \left( 1-f_{QCD}^W \right) - 
\epsilon_W  N_{obs}^Z (1-f_{QCD}^Z) 
\frac{A_{Zee}^W + A_{Z\tau}^W}
{A_Z  \epsilon_Z} }
{
\epsilon_W  A_{W} 
\left( 1+\frac{A_{W\tau}^W}{A_W} \right)
 {\cal L}}\]
where $N_{obs}^W$ and $N_{obs}^Z$ are the number of \Wev\ and \Zee\ candidate 
events; 
$f_{QCD}^W$ and $f_{QCD}^Z$ are the fraction of the \Wev\ and \Zee\ candidate 
events that come from
multijet, $b$ quark, and direct photon background sources; 
$\epsilon_W$ and $\epsilon_Z$ are the efficiency for \Wev\ and \Zee\ events
to pass the selection requirements; 
$A_W$ and $A_Z$ are the geometric and kinematic acceptance for \Wev\ and 
\Zee\ which include effects from detector resolution; 
$A_{W\tau}^W$, $A_{Zee}^W$ and $A_{Z\tau}^W$ are the fraction of \Wtv\, \Zee\, 
and \Ztt\ events that passes the \Wev\ selection 
criteria;
and $\cal L$ is the integrated luminosity of the data sample.  

The product of the $Z$ boson cross section and the branching fraction for 
\Zee\ is determined from the relation
\[
\sigma(p\overline{p} \rightarrow Z+X)\cdot B(Z \rightarrow ee) =\]
\[
\frac{N_{obs}^Z  \left( 1 - f_{QCD}^Z \right) 
\left( 1 -f_{DY}\right)
}{\epsilon_Z  A_Z  {\cal L}}\]
where $f_{DY}$ is a correction for the Drell-Yan contribution to
$Z$ boson production. The results are summarized in table~\ref{tab:d0xsec}.

 The ratio of
the cross sections can be used to extract an indirect measurement of the
total width of the $W$ boson. In
the ratio, many of the systematic uncertainties, including the
luminosity uncertainty, cancel. This method therefore gives the most
precise determination of the $W$ width currently available.
The result on the cross section ratio is summarized in table~\ref{tab:d0xsec}.
Using this results we can determine the
electronic branching fraction of the $W$ boson via
$ B( W \rightarrow e \nu ) = {\cal R} B(Z \rightarrow ee)
\frac{\sigma_Z}{\sigma_W}$.
Using $B( Z \rightarrow ee)$ = 0.03367 $\pm$ 0.00006~\cite{lepzee} and
$\sigma_W / \sigma_Z$ = 3.29 $\pm$ 0.03~\cite{theoryrs}, we get
$B(W\rightarrow e \nu)$ = \brwev\ $\pm$ \brstat\ (stat)
$\pm$ \brsyst\ (syst) $\pm$ \brthy\ (other) $\pm$ \brnlo\ (NLO), where the 
next-to-last source of 
uncertainty comes from uncertainties in $B( Z \rightarrow ee)$ and 
in $\sigma_W / \sigma_Z$.
Assuming  the standard model prediction for the electronic
partial width ($0.2270 \pm 0.0011\;\rm GeV$~\cite{wwidth}), 
we can calculate the $W$ boson width
$\Gamma_W = \Gamma_W^e / B(W \rightarrow e \nu)$ as
\gw\ $\pm$ \gwstat\ (stat) $\pm$ \gwsyst\ (syst) $\pm$ \gwthy\ (other) 
$\pm$ \gwnlo\ (NLO) GeV, 
to be compared with the standard model prediction of
$\Gamma_W$ = 2.094 $\pm 0.006\;\rm GeV$~\cite{wwidth}.
The difference between our measured value and the standard model prediction, 
which is the width for the $W$ boson to decay to final states
other than the two lightest quark doublets and the three
lepton doublets, is thus 0.036 $\pm 0.060\;\rm GeV$. This is
consistent with zero within uncertainties, so we set a $95\%$ confidence level 
upper limit on the $W$ boson width to non-standard-model final states 
(``invisible width") of \gwinvgev\ GeV.

\begin{figure}[h]
\vspace{9pt}
\vskip-1.2cm
\psfig{figure=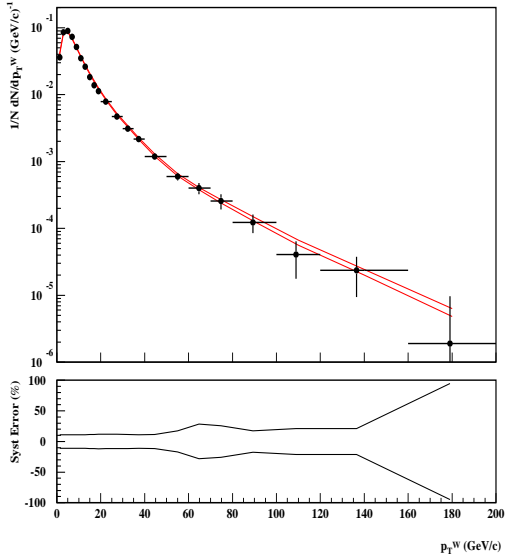,height=8.5cm,width=7.3cm}
\vskip-1cm
\caption{The $W$ boson transverse momentum spectrum, showing the
D\O\ result (solid points) with statistical uncertainty.
The theoretical calculation by Arnold and Kauffman~\protect\cite{ak}, 
smeared for detector
resolutions, is shown as two lines corresponding to the 
$\pm 1\sigma$ variations of the uncertainties in the detector modeling.
The fractional systematic uncertainty on the data 
is shown as a band in the lower portion of the plot.}
\label{fig:d0ptw}
\end{figure}

\begin{figure}[t]
\vspace{9pt}
\vskip-0.5cm
\psfig{figure=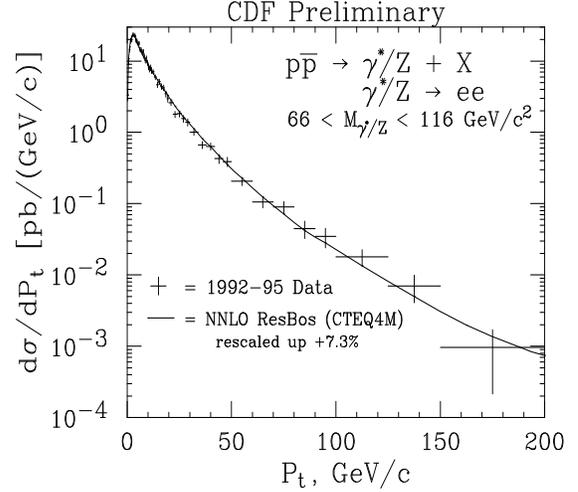,height=6.5cm,width=7.3cm}
\caption{The $Z$ boson transverse momentum spectrum, showing the
CDF result (solid points) with total uncertainty. The data have been
unfolded for detector resolutions and are compared to the theoretical
calculation by Balasz and Yuan~\protect\cite{ly}, 
scaled up $7.3\%$ to match the
measured inclusive $Z$ production cross section. }
\label{fig:cdfzpt}
\end{figure}

\begin{figure}[t]
\vspace{9pt}
\psfig{figure=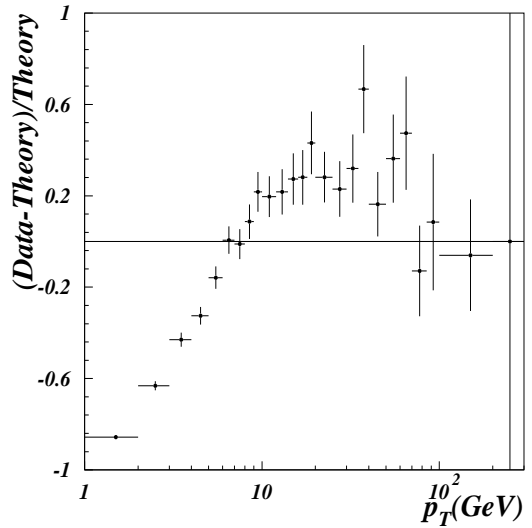,height=7.6cm,width=7.3cm}
\vskip-1cm
\caption{Fractional difference between \D0 $Z$ transverse momentum
data and the fixed-order calculation from Arnold and Reno~\protect\cite{ar}.}
\label{fig:dylan}
\end{figure}

CDF and \D0 measure  the differential $d\sigma/dp_T$ distribution for $W$ and
$Z$ bosons decaying to electrons. The data agrees with the combined QCD 
perturbative and resummation calculations~\cite{ak,ly}, as can be seen in 
figure~\ref{fig:d0ptw} for the \D0 $W$ data, and in figure~\ref{fig:cdfzpt}
for the CDF $Z$ data. In addition, the $d\sigma/dp_T$
 distribution for the $Z$ boson discriminates
between different vector boson production models and can be used to
extract values of the non-perturbative parameters for the resummed
prediction from a fit to the differential cross section.
Figure~\ref{fig:dylan} compares \D0 $Z$ 
data to the fixed-order perturbative QCD theory~\cite{ar}
 in terms of a percentage difference from the
prediction. We observe a strong disagreement at low-\pt, as expected
due to the divergence of the NLO calculation at \pt$=0$, and a
significant enhancement of the cross section relative to the
prediction at moderate values of \pt, confirming the enhancement of
the cross section from soft gluon emission.

\clearpage

\begin{table*}[hbt]
\setlength{\tabcolsep}{1.5pc}
\newlength{\digitwidth} \settowidth{\digitwidth}{\rm 0}
\catcode`?=\active \def?{\kern\digitwidth}
\caption{The \D0 preliminary cross sections for $W$ and $Z$ bosons and
their ratio.}
\label{tab:d0xsec}
\begin{tabular*}{\textwidth}{@{}l@{\extracolsep{\fill}}ccc}
\hline
           & Value & Uncertainty Contribution(pb) \\ 
\hline
${N_{obs}^W}$  & \wnum                 & 10   \\
${\epsilon_W}$ & \weff\ $\pm$ \wefferr & 30  \\
${A_W}$ & \wacc\ $\pm$ \waccerr        & 20  \\
$f_{QCD}^W$    & \wfqcd\ $\pm$ \wfqcderr  & 35  \\
$(A_{Zee}^W + A_{Z\tau}^W)/A_Z$ & \wacczinw\ $\pm$ \wacczinwerr & -- \\
${\epsilon_Z}$ & \zeff\ $\pm$ \zefferr    & --  \\
$f_{QCD}^Z$    & \zfqcd\ $\pm$ \zfqcderr  & --   \\
$N_Z^W$        & $621 \pm 155$            & 6    \\
$A_{W\tau}^W/A_W$ & 0.0211 $\pm$ 0.0021   & 5   \\ 
$\cal L$          & \lumb\ $\pm$ \lumberr\ \ipb & 100 \\
\hline\hline
& &  \\
$\sigma(p \overline{p} \rightarrow W + X) \cdot B(W \rightarrow e \nu)$   
& \wxsec\ $\pm 10{\rm(stat)} \pm 50 {\rm(syst)} \pm 100{\rm(lum)}$ pb & \\
& & \\
\hline\hline
${N_{obs}^Z}$  & \znum  &  3  \\
${\epsilon_Z}$ & \zeff\ $\pm$ \zefferr & 3  \\
${A_Z}$        & \zacc\ $\pm$ \zaccerr & 2  \\
$f_{QCD}^Z$    & \zfqcd\ $\pm$ \zfqcderr & 1    \\
$f_{DY}$       & \zfdy\ $\pm$ \zfdyerr & $< 1$  \\
$\cal L$       & \lumb\ $\pm$ \lumberr\ \ipb & 10 \\
\hline\hline
& & \\
$\sigma(p \overline{p} \rightarrow Z + X) \cdot B(Z \rightarrow ee)$
      & \zxsec\ $\pm 3{\rm(stat)} \pm 4 {\rm(syst)} \pm 10{\rm(lum)}$ pb & \\
& & \\
\hline\hline
 $N_{obs}^W/N_{obs}^Z$ & \rnum\ $\pm$ \rnumerr &  0.15\\
 $\epsilon_Z/\epsilon_W$ & \reff\ $\pm$ \refferr & 0.06 \\
 $A_{Z}/A_W$ & \racc\ $\pm$ \raccerr &  0.09\\
 $(A_{Zee}^{W}+A_{Z\tau}^{W})/A_Z$ & \wacczinw\ $\pm$ \wacczinwerr & 0.03\\
 $f_{QCD}^W$ & \wfqcd\ $\pm$ \wfqcderr & 0.16\\
 $f_{QCD}^Z$ & \zfqcd\ $\pm$ \zfqcderr & 0.05 \\
 $f_{DY}$ & \zfdy\ $\pm$ \zfdyerr & 0.01\\
 $A_{W\tau}^W/A_W$ & \wacctau\ $\pm$ \wacctauerr & 0.02\\ 
\hline\hline
 & & \\
 $\cal R$      & \rxsec\  $\pm 0.15{\rm(stat)} \pm 0.20 {\rm(syst)} 
\pm 0.10{\rm(NLO)}$& \\
               &                     & \\
\hline\hline
\end{tabular*}
\end{table*}
\clearpage

\subsection{Photon Production}

\begin{figure}[htb]
\vspace{9pt}
\vskip0.8cm
\hskip-1.5cm
\psfig{figure=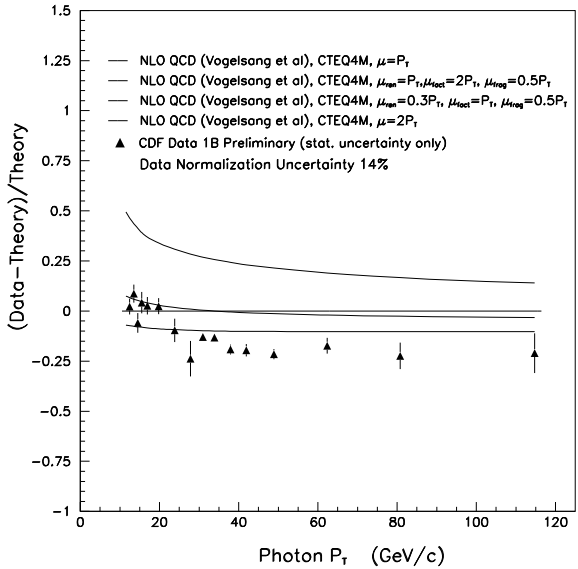,height=8.5cm,width=8.5cm}
\vskip-2cm
\caption{Comparison of the CDF inclusive photon cross section with the
latest QCD predictions~[20] shown using a wide range of scale choices.}
\label{fig:cdfphoton}
\end{figure}

CDF measures the inclusive photon cross section~\cite{Kuhlmann}
in the central region
$|\eta|<0.9$ using $87\;\rm pb^{-1}$ taken during the 1994--1995 
\ppbar collider run. Figure~\ref{fig:cdfphoton} shows the data compared
to variations of the model by Vogelsang \etal \cite{vogel}, in which
the renormalization, fragmentation and factorization scales are changed
independently. None of these changes allows the theory to agree with the data
 over the entire $E_T$ region.

CDF also measures the production of a photon plus a muon~\cite{Kuhlmann}
 and probes the
charm content of the proton via the reaction 
$c g \to c \gamma \to \gamma \mu X$.
Results for the production cross section are compared with a NLO calculation
and to {\sc PYTHIA}~\cite{pythia} Monte Carlo. 
The latter, which does not include bremsstrahlung 
radiation, falls below the data; the data are consistent with the NLO
prediction.

\begin{figure}[htb]
\vspace{9pt}
\psfig{figure=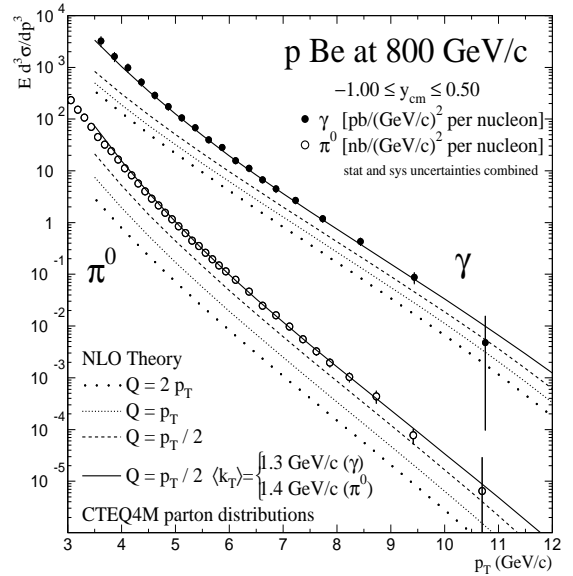,height=7.6cm,width=7.3cm}
\caption{Invariant cross sections for direct--$\gamma$ and $\pi^0$
production from E706. Curves represent the NLO pQCD prediction with and
without supplemental parton $k_T$ smearing.}
\label{fig:e706}
\end{figure}

E706~\cite{begel} uses data accumulated from a proton beam at 
 $800\;\rm GeV/{\it c}$ on Be target and 
measures the $\pi^0$ and direct--photon
inclusive cross section as functions of $p_T$. The measurements are shown in
figure~\ref{fig:e706} compared to NLO QCD with and without $k_T$ 
enhancement~\cite{e706th}. 
Current pQCD calculations fail to account for the measured cross sections using
conventional choices of scales. A simple implementation of supplemental parton
$k_T$ in pQCD calculations~\cite{e706LO}, with $<k_T> \sim 1$, 
provides a reasonable description of the data.
E706 obtained similar results using a proton beam at $530\;\rm GeV/{\it c}$,
and a $\pi^{-}$ beam at $515\;\rm GeV/{\it c}$, and using a hydrogen target.

\subsection{Drell--Yan production}

\begin{figure}[htb]
\vspace{9pt}
\hskip-2.5cm
\psfig{figure=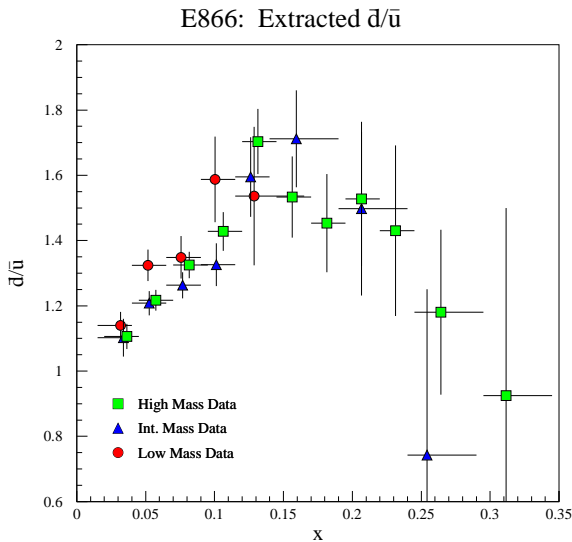,height=9.5cm,width=9.5cm}
\vskip-4cm
\caption{The ratio  $\overline{d}/\overline{u}$ in the proton as a function
of $x$ extracted from the E866 cross section ratios.}
\label{fig:e866}
\end{figure}

E866 measures the Drell--Yan cross 
sections to muon pairs~\cite{isenhower} with dimuon mass
$M_{\mu^{+}\mu^{-}} \ge 4.5\;\rm GeV/{\it c}$ from an  
$800\;\rm GeV/{\it c}$ proton beam incident on hydrogen and deuterium targets.
The ratio of $\overline{d}/\overline{u}$ in the proton as a
function of Bjorken $x$ is determined 
from the ratio of Drell--Yan cross sections. 
The result is shown in figure~\ref{fig:e866}. The ratio 
$\overline{d}/\overline{u}$ decreases with $x$ for $x>0.2$. 
However, at moderate $x$, $0.05 < x < 0.2$, it clearly is different from
one and only approaches unity as $x \to 0.025$.

\D0 measures the Drell--Yan cross section in the dielectron invariant mass
range from 50 to $1000\;\rm GeV/{\it c}^2$ using $120\;\rm pb^{-1}$ of data
collected in \ppbar collisions at $\sqrt{s}=1.8\;\rm TeV$. No deviation
from the standard model expectations is observed, and the data are used to
set limits on the energy scale of quark--electron compositeness with
common constituents. The $95\%$ confidence level lower limits on the 
compositeness scale vary between $3.3\;\rm TeV$ and $6.1\;\rm TeV$ depending
on the assumed form of the effective contact interaction~\cite{d0dy}.

\section{MEASUREMENT OF THE $W$ MASS}
In the standard model of the electroweak interactions, the mass
of the $W$ boson is predicted to be
\begin{equation}
\mw = \left( \frac{\pi\alpha(\mz^2)}
 {\sqrt{2}G_F}\right)^\frac{1}{2} \frac{1}{\sin\theta_w\sqrt{1-\Delta r_W}}\; .
\label{eq:dr}
\end{equation}
In the ``on-shell'' scheme~\cite{on_shell} $\cos\theta_w = \mw/\mz$,
where $\mz$ is the $Z$ boson mass.  A measurement of $\mw$,
together with $\mz$, the Fermi constant ($G_F$), and the electromagnetic
coupling constant ($\alpha$),
determines the electroweak radiative corrections $\Delta r_W$ experimentally.
Purely electromagnetic corrections are absorbed into the value of $\alpha$
by evaluating it at $Q^2=\mz^2$~\cite{amz}.
The dominant contributions to $\Delta r_W$
arise from loop diagrams that involve the top quark and the Higgs boson.
The correction from the $t\overline b$ loop is substantial
because of the large mass difference between the two quarks. It is proportional
to $m_t^2$ for large values of the top quark mass $m_t$. Since $m_t$ has
been measured at the Tevatron, this
contribution can be calculated within the Standard Model. For a large Higgs
boson mass, $m_H$, the correction from the Higgs loop is proportional to $\ln
m_H$. 
If additional particles which couple to the $W$ boson exist, they would give
rise to additional contributions to $\Delta r_W$. Therefore, a measurement of
$\mw$ is one of the most stringent experimental tests of SM predictions.
Deviations from the predictions may indicate the existence of new physics.
Within the SM, measurements of $\mw$ and the mass of the top quark constrain
the mass of the Higgs boson. A discrepancy with the range allowed by the 
Standard Model could indicate new physics.
The experimental challenge is thus to measure the $W$
boson mass to sufficient precision, about $0.1\%$, to be sensitive to these
corrections.

\begin{figure}[t]
\vspace{9pt}
\vskip-1.5cm
\hskip-1.1cm
\psfig{figure=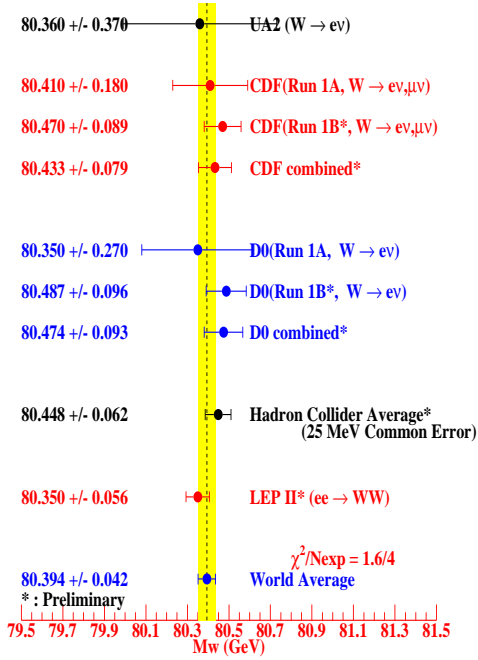,height=13.0cm,width=8.0cm}
\vskip-3cm
\caption{Summary of current direct measurements of the $W$ mass. The world
average is shown as the band.}
\label{fig:wmass}
\end{figure}

\begin{figure}[t]
\vspace{9pt}
\vskip-2cm
\psfig{figure=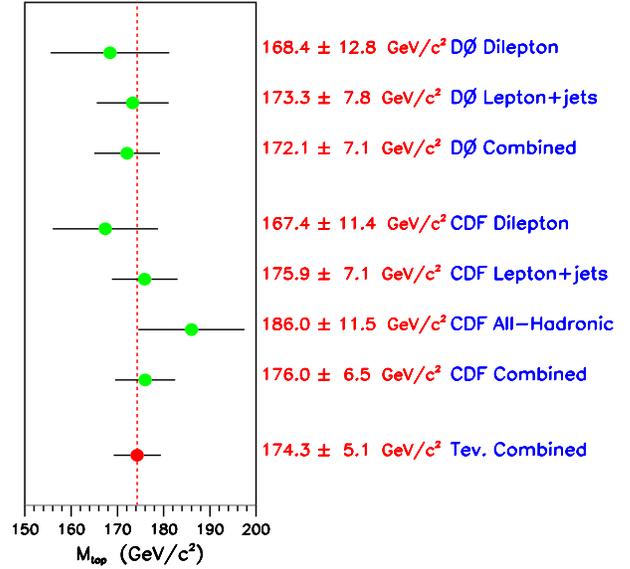,height=8.0cm,width=8.0cm}
\caption{Summary of current measurements of the top quark mass from the 
Fermilab collider experiments CDF and \D0. The combined result is shown as
a line.}
\label{fig:mtop}
\end{figure}

\begin{figure}[b]
\vspace{9pt}
\vskip-5cm
\psfig{figure=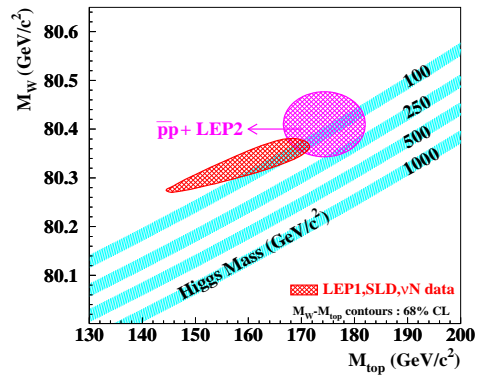,height=10.0cm,width=8.0cm}
\vskip-3cm
\caption{Current world averages for the direct $W$ boson and top quark mass 
measurements from \D0, CDF and LEP experiments. The bands show SM predictions
for the indicated Higgs masses~\protect\cite{mw_v_mt}.}
\label{fig:higgs}
\end{figure}

CDF and \D0 report precise new measurements of the $W$ boson mass based on
an integrated luminosity of $\sim 100\;\rm pb^{-1}$ from \ppbar collisions at
$\sqrt{s}=1.8$~TeV. CDF identifies $W$ bosons by their decays to $e\nu$ 
and $\mu\nu$, with electrons and muons identified in the central pseudorapidity
region of the detector. The combined CDF result~\cite{cdfmw} for $\mw$ is
$\mw({\rm CDF})= 80.433\pm0.079\;\rm GeV$. 
\D0 identifies $W$ bosons by their decays 
to $e\nu$, with electrons identified both in the central and the forward
pseudorapidity region.  The combined \D0 result~\cite{d0mw} for $\mw$ is
$\mw({\rm \D0})=80.474\pm0.093\;\rm GeV$. 
The Tevatron values are combined with a 25 MeV common error, covering 
common uncertainties in pdf's, W width and QED
corrections to give : 
$\mw({\rm Tevatron})=80.450\pm0.063\;\rm GeV$.
The inclusion of the UA2 data produces a hadron collider average of : 
$\mw(\pbarp)=80.448\pm0.062\;\rm GeV$. 
Including LEP2 gives a world average of direct W mass measurements of : 
$\mw({\rm direct})=80.410\pm0.044\;\rm GeV$. The various measurements can be
seen in figure~\ref{fig:wmass}. Figure~\ref{fig:mtop} shows the current
results for the top quark mass measurements from CDF and \D0.  
Figure~\ref{fig:higgs} compares the direct measurements of the $W$ boson and
top quark masses to
the values predicted by the Standard Model for a range of Higgs mass values
\cite{mw_v_mt}.The measured values are in agreement with the prediction
of the Standard Model.

\newpage

\clearpage

\section{STUDIES OF $CP$ VIOLATION}

CDF reports an updated~\cite{CDFpub} 
direct measurement of the Standard Model CP violation
parameter ${\rm sin}2\beta$ using $110\;\rm pb^{-1}$ of data collected in
\ppbar collisions at $\sqrt{s}=1.8\;\rm TeV$.
CP violation can manifest itself as an asymmetry in the decay rate of
particle versus antiparticle to a particular final state:
\[
A_{CP}=\frac{N(\overline{B^0}  \to J/\psi K_S^0) - N(B^0 \to J/\psi K_S^0)}
{N(\overline{B^0}  \to J/\psi K_S^0) + N(B^0 \to J/\psi K_S^0)}.
\]
In the Standard Model, this CP asymmetry is proportional to ${\rm sin}2\beta$.
A value of ${\rm sin}2\beta > 0$ would indicate CP violation in the $b$
quark system. 

\begin{figure}[htb]
\vspace{9pt}
\psfig{figure=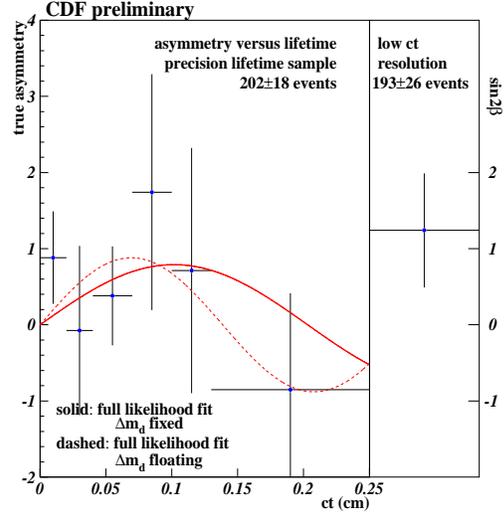,height=7.6cm,width=7.3cm}
\caption{The true asymmetry (${\rm sin}2\beta$) as a function of lifetime for
$B \to J/\Psi K_S^0$ events from CDF. The non--SVX events that lack
precision lifetime information are shown as a single point on the right.}
\label{fig:cdfcp}
\end{figure}

CDF uses a signal sample that consists of 
$~400\;\; B \to J/\psi K_S^0$ events.
 About half of the events have both
muon tracks fully contained within the silicon vertex detector, and 
therefore have precision lifetime information. Three tagging methods are used
to identify the flavor of the $B$ meson at the time of 
production~\cite{CDFtag}:
the same--side tag, the soft--lepton tag and a jet--charge tag. 
Figure~\ref{fig:cdfcp} shows the true asymmetry (${\rm sin}2\beta$) as a 
function of lifetime for $B \to J/\psi K_S^0$ events. The non--SVX events, 
for which precision lifetime information is not available, are shown
as a single point on the right. A maximum likelihood method is used
to extract the result of ${\rm sin}2\beta = 0.79 \pm 0.39 {\rm (stat)}
\pm 0.16{\rm (syst)}$. This measurement is the best indication that $CP$
symmetry is violated in the $b$ quark system and is consistent with the 
Standard Model expectation of a large positive value of 
${\rm sin}2\beta$~\cite{cpth}. 

KTeV reports a new measurement of the direct $CP$ violation parameter
$Re(\epsilon'/\epsilon)$ using $23\%$ of the data collected during the
1996--1997 fixed target run. It is well known that, 
in the neutral $K$ meson system, the two strangeness states 
$(K^{0},\overline{K}^{0})$ mix to produce short and long lived kaons
$(K_{S},K_{L})$. In 1964, the observation of $K_L \to \pi \pi$ 
decays~\cite{ktev1} revealed 
that $CP$ symmetry is violated by the weak interaction. The dominant effect 
is an asymmetry in the $K^{0} - \overline{K}^{0}$ mixing, parameterized by 
$\epsilon$. Direct $CP$ violation refers to the
$K^{0}_{S,L} \to \pi^{+} \pi^{-}$ decays, in which a true $CP$ eigenstate 
decays to a final state with opposite $CP$. The most sensitive searches 
for direct $CP$ violation~\cite{ktev2} measure
\[ Re(\epsilon'/\epsilon) \approx \frac{1}{6}\left [ 
\frac{\Gamma(K^{0}_{L} \to \pi^{+} \pi^{-})/
\Gamma(K^{0}_{S} \to \pi^{+} \pi^{-})}{
\Gamma(K^{0}_{L} \to \pi^{0} \pi^{0})/ 
\Gamma(K^{0}_{S} \to \pi^{0} \pi^{0})} - 1 \right ]. \]

In this experiment, $800\;\rm GeV$
 protons from the Tevatron are used to produce two $K^{0}_{L}$ beams; 
a regenerator in one of the beams (alternating once per minute) converts 
some $K^{0}_{L}$ to $K^{0}_{S}$ by coherent forward scattering. Kaon 
decays in the ``vacuum beam'' ($K^{0}_{L}$) and ``regenerator beam'' 
(mostly $K^{0}_{S}$) are collected simultaneously with the KTeV detector.
 $Re(\epsilon'/\epsilon)$ is extracted from the data using a fitting program 
which calculates decay distributions using full treatments of kaon 
production and regeneration. The measured value of
$Re(\epsilon'/\epsilon)=(28.0\pm3.0({\rm stat})\pm2.8({\rm syst}))
\times 10^{-4}$ firmly establishes the existence of $CP$ violation in the 
decay process and rules out the ``superweak'' model~\cite{ktevth1}. 
Standard Model calculations of $Re(\epsilon'/\epsilon)$ depend sensitively
on input parameters and the method of calculations~\cite{ktevth2}; it 
remains to be seen whether this large value of $Re(\epsilon'/\epsilon)$
can be accommodated or may be an indication of new physics beyond the
Standard Model.

\section{CONCLUSIONS}

Although the Tevatron experiments have stopped taking data several years ago,
the number of new results is overwhelming. The unprecedented precision in the
experimental results that is being achieved is confronting theory
with experiments at new limits. So far, QCD has held up to all the
quantitative tests that were performed. We expect to see improvements in the
calculations in the following years while the experiments prepare for a new
period of data taking in which the Tevatron will continue to improve our 
understanding of nature.

\section{ACKNOWLEDGMENTS}

I would like to thank my Tevatron colleagues who have provided me with
the results included in this review, especially Andrew Brandt, Daniel
Elvira, Ulrich Heintz, Rob Snihur and Mike Strauss from \D0,
Alex Akopian, Mike Albrow, Frank Chlebana and Barry Wicklund from CDF,
Marek Zielinski from E706, Donald Isenhower from
E866,  Peter Shawhan from KTeV, and Jeff Appel from E791.
I would also like to thank the DIS99 organizers for an extremely interesting
workshop.

\end{document}